\documentclass[12pt]{article}
\pdfoutput=1%permet de dire que les figures sont des .pdf (pour JHEP)
\usepackage[square,comma,numbers,sort&compress]{natbib}
\usepackage{graphicx,epstopdf,amssymb,amsfonts,amsmath,amsthm,array,
mathrsfs,amscd,enumitem}%,wick}
\usepackage{float}
\usepackage{xcolor}
\usepackage{appendix}
\usepackage{fnpct}%pour des notes en bas de page multiples
\usepackage{lmodern}
\usepackage[T1]{fontenc}
\usepackage[utf8]{inputenc}
\usepackage{hyperref}% hyperliens dans le fichier .pdf; toujours charger ce package en dernier!!!
\newcommand{\pbs}[1]{\let\temp=\\#1\let\\=\temp}
\numberwithin{equation}{section}
\def\be{\begin{equation}}\def\ee{\end{equation}}
\def\cvp{\raise 2pt\hbox{,}} 
 
\def\im{\mathop{\text{Im}}\nolimits}

 \def\d{{\rm d}} 
\def\la{\lambda}\def\La{\Lambda}

\DeclareMathOperator{\diff}{Diff}\DeclareMathOperator{\met}{Met}

\DeclareMathOperator{\Hilb}{\text{Hilb}}

\def\disk{\mathscr D}

\def\diffSp{\diff_{+}(\Sone)}

\def\Sone{\text{S}^{1}}

\def\PslR{\text{PSL}(2,\mathbb R)}

\def\PslR{\text{PSL}(2,\mathbb R)}

\def\la{\lambda}
\def\La{\Lambda}

\def\disk{\mathscr D}

\def\met{\text{Met}}

\def\plb#1#2#3{{\it Phys.\ Lett.\ }{\bf B #1} (#2) #3}
\def\npb#1#2#3{{\it Nucl.\ Phys.\ }{\bf B #1} (#2) #3}

\def\prl#1#2#3{{\it Phys.\ Rev.\ Lett.\ }{\bf #1} (#2) #3}
\def\jhep#1#2#3{{\it J. High Energy Phys.\ }{\bf #1} (#2) #3}
\def\prd#1#2#3{{\it Phys.\ Rev.\ }{\bf D #1} (#2) #3}

\def\cmp#1#2#3{{\it Comm.\ Math.\ Phys.\ }{\bf #1} (#2) #3}

\def\mpla#1#2#3{{\it Mod.\ Phys.\ Lett.\ }{\bf A #1} (#2) #3}

\def\rmp#1#2#3{{\it Rev.\ Mod. Phys. }{\bf #1} (#2) #3}

\def\imath#1#2#3{{\it Invent math }{\bf #1} (#2) #3}

\def\jpa#1#2#3{{\it J.\ Phys.\ }{\bf A #1} (#2) #3}

\def\ahpa#1#2#3{{\it Ann.\ I.~H.~Poincar\'e }{\bf A #1} (#2) #3}

\begin{document}

{\pagestyle{empty}
\parskip 0in
\

\vfill
\begin{center}
%{\sffamily\large\bfseries SPECTRAL ASYMMETRY AND SUPERSYMMETRY}
{\LARGE Jackiw-Teitelboim Gravity,}

\bigskip

{\LARGE Random Disks of Constant Curvature,}

\bigskip

{\LARGE Self-Overlapping Curves and Liouville $\text{CFT}_{1}$ }

%\bigskip

%{\LARGE\sffamily at Infinite and Finite Cutoff}

\vspace{0.4in}

%Frank F{\scshape errari}%{\renewcommand{\thefootnote}{$\!\!\dagger$}
%\footnote{On leave of absence from Centre National de la Recherche
%Scientifique, Laboratoire de Physique Th\'eorique de l'\'Ecole Normale
%Sup\'erieure, Paris, France.}}

Frank F{\scshape errari}

\medskip
{\it Service de Physique Th\'eorique et Math\'ematique\\
Universit\'e Libre de Bruxelles (ULB) and International Solvay Institutes\\
Campus de la Plaine, CP 231, B-1050 Bruxelles, Belgique}

%\medskip
%
%{\it Okinawa Institute Of Science and Technology\\
%Okinawa, Onna 904-0495, Japan
%}

\smallskip
{\tt frank.ferrari@ulb.be}
\end{center}
\vfill\noindent

We propose a microscopic definition of finite cut-off JT quantum gravity on the disk, both in the discretized and in the continuum points of view. The discretized formulation involves a new model of so-called self-overlapping random polygons. The measure is not uniform, implying that the degrees of freedom are not in one-to-one correspondence with the shape of the boundary. The continuum formulation is based on a boundary $\text{CFT}_{1}$ from which we predict some critical exponents of the self-overlapping polygon model. The coupling to an arbitrary bulk matter CFT is also discussed.

\vfill

%\noindent Gravit\'e, Th\'eories de Jauge et Cordes,}
%Les Houches summer school 2001, Session LXXVI.
\medskip
%
%\vfill
\begin{flushleft}
\today
\end{flushleft}
\newpage\pagestyle{plain}
\baselineskip 16pt
\setcounter{footnote}{0}

\section{Introduction}

Jackiw-Teitelboim quantum gravity \cite{JTpapers} is a two-dimensional quantum gravitational theory in which the bulk curvature is fixed, but the extrinsic curvature of the boundary is allowed to fluctuate. Restricting to the case of disk topology, the partition function can be formally written as a path integral 
\be\label{ZJTgen} \mathscr Z^{(\eta)}(\ell,\La) = \int_{\met^{(\eta)}_{\ell}(\disk)}e^{-\frac{\La}{16\pi}A[g]} \, \d\mu (g)\ee
over the space of constant curvature $R=2\eta$ metrics $g$ of fixed boundary length $\ell$ on the disk $\disk$, with $\eta = -1$, $0$ or $+1$. We note $A[g]$ the area functional, $\La$ the cosmological constant and $\d\mu(g)$ an appropriate diffeomorphism-invariant integration measure. The path integral \eqref{ZJTgen} is traditionally written in terms of a dilaton field $\Phi$ that plays the role of a Lagrange multiplier enforcing the constraint $R=2\eta$; for $\eta=\pm 1$ the cosmological constant is then proportional to the boundary value of $\Phi$ according to $\La = 2\eta\Phi_{|\partial\disk}$.

%The classical action is usually written in terms of the metric $g$, Ricci scalar $R$, boundary extrinsic curvature and arc-length coordinate $k$ and $s$, and dilaton field $\Phi$, as
%%
%\begin{multline}\label{Scl} S = -\frac{1}{16\pi}\biggl[\int_{\disk}\Phi\bigl(R - 2\eta\bigr)\sqrt{g}\,\d^{2}x + 2 \oint_{\partial\disk} \Phi k\, \d s \\+ \bigl(1-|\eta|\bigr)\frac{\La}{16\pi}\int_{\disk}\sqrt{g}\,\d^{2}x \biggr]\, .
%\end{multline}
%%
%The dilaton $\Phi$ plays the role of a Lagrange multiplier field imposing the constraint $R=2\eta$ in the bulk. The parameter $\eta$ is chosen to be $-1$, $0$ or $+1$, corresponding to the three possible versions of the theory, in negative, zero or positive curvature. The boundary conditions are chosen such that the boundary value of $\Phi$ and the boundary length are fixed,
%%
%\be\label{bdcond} \Phi_{|\partial\disk} = \varphi = \frac{1}{2}\eta\La\, ,\quad \oint_{\partial\disk}\d s  = \ell\, .\ee
%%

The JT models have attracted considerable interest in recent years, due to relations with the physics of near-extremal black holes, holography, SYK and tensor models, etc., see e.g.\ \cite{JTappli1,JTappli2,JTappli3}. The vast majority of studies have focused on the negative curvature case in the near-AdS limit $\ell\rightarrow\infty$, $\La\rightarrow -\infty$, $\ell/|\La|$ fixed, where smooth wiggly boundary configurations, described by the so-called ``reparameterization ansatz'' and governed by the Schwarzian action, are postulated to dominate \cite{JTappli1}, see Fig.\ \ref{fig1}.

The present work is motivated by the observation that, in spite of the
many important applications, the theoretical foundations of JT quantum
gravity are very poorly understood compared to the one achieved by standard two-dimensional Liouville gravity \cite{Liouvilleref1,Liouvilleref2}, which is by now a part of mathematics \cite{Liouvilleref3}. In spite of some rare efforts \cite{Micpapers1,Micpapers2,Micpapers3}, a proper microscopic, UV-complete, definition of the models has not been proposed so far and the meaning of the formal path integral \eqref{ZJTgen} remains elusive, especially for finite values of $\ell$ (the ``cut-off''\footnote{The parameter $\ell$ governs the size of the geometry and is thus an IR cut-off from the gravity point of view. In negative curvature JT it also plays the role of a UV cut-off for the putative holographic dual. In Section \ref{SOPSec} we also introduce a UV cut-off $\ell_{0}$ in the gravitational theory, associated with a lattice discretization.}) and $\La$ in negative curvature or in general for the zero and positive curvature models, which are of considerable potential interest, for instance for applications in cosmology. Let us emphasize that the UV structure does not depend on large scale effects like the choice of bulk curvature or topology. Our main results and ideas are thus completely general, but we focus on the disk to make the discussion technically simpler. 

We propose below two approaches, which we conjecture to be equivalent, to JT quantum gravity.  One approach is based on a new random polygon model which describes the discretized fluctuating disk boundary. The allowed polygons are so-called self-overlapping curves (SOPs) that must be counted with a non-trivial multiplicity. The multiplicity implies that the fundamental metric degrees of freedom of the theory are \emph{not} in one-to-one correspondence with the shape of the boundary. We postulate that the continuum limit of the SOP model yields JT quantum gravity on the disk. Another approach involves working directly in the continuum. We show that it is possible to adapt the techniques used in Liouville theory to obtain a formulation of JT gravity coupled to an arbitrary bulk matter CFT of central charge $c$. The fundamental degrees of freedom are encoded in the boundary Liouville field. In the $\eta=0$ flat case, the boundary action takes an elegant Hilbert-Liouville form for a log-correlated boundary field, from which we extract exact critical exponents. We find that $c=0$  is a barrier, in the same way as $c=1$ is a barrier in Liouville theory. The geometry becomes semiclassical when $c\rightarrow -\infty$, whereas a proliferation of baby universes is expected for $c>0$, see Fig.\ \ref{fig1}. These ideas will be presented in much greater detail in forthcoming publications \cite{companion1,companion2,more1,more2}.

\begin{figure}
\centerline{\includegraphics[width=6in]{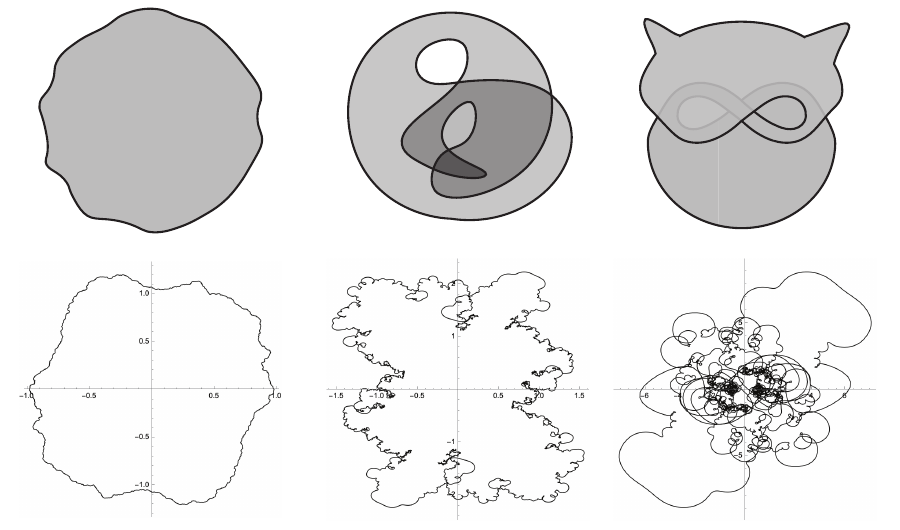}}
\caption{\label{fig1} Examples of distorted disks. Upper-left: a smoothly wiggling boundary described by the reparameterization ansatz; Upper-center: an immersion with overlaps; Upper-right: the Milnor example with multiplicity two; Lower insets, from left to right: typical self-overlapping curves from the Hilbert-Liouville model, Eq.\ \eqref{HilbLiouaction}, for $Q^{2}=25$ (semi-classical regime), $Q^{2}=4$ (pure gravity) and $Q^{2}=2$ (exploding universe).}
\end{figure}

\section{\label{SOPSec}Random self-overlapping polygons}

Constant curvature metrics on the disk can be described by immersions $F:\disk\rightarrow \mathbb T_{\eta}$ from $\disk$ into a canonical target space, which is the Euclidean plane, the hyperbolic space or the round two-sphere for $\eta=0$, $-1$ or $+1$, respectively. The metric is obtained by pullback, $g=F^{*}\delta$, where $\delta$ is the canonical metric on $\mathbb T_{\eta}$. The image of the source disk $\disk$ is called a distorted disk. It is a convenient, isometric representation of the disk endowed with a metric. Various examples are depicted in  Fig.\ \ref{fig1}. When the boundary curve of the distorted disk has no self-intersection (self-avoiding loop), the distorted disk has no overlap and the associated immersion $F$ is an embedding. The configurations described by the reparameterization ansatz (upper-left inset in the figure) and more general cases, that have been considered in \cite{Micpapers2, Micpapers3} and for which zig-zags are allowed, are in this self-avoiding class. It is crucial to understand that these examples are very atypical. Descriptions limited to such configurations are at best very rough, similar to minisuperspace approximations. \emph{For a typical configuration, the boundary curve has self-intersections, the immersion $F$ is not an embedding and the distorted disk has overlaps, see Fig.\ \ref{fig1}.}

Note, however, that an arbitrary closed curve with self-intersections is not permitted, because it does not bound a distorted disk and thus is not associated with a metric. A theory based on arbitrary closed random loops (Brownian bridges, boundary quantum particle description), as proposed in \cite{Micpapers1}, is thus not a theory of quantum gravity.

Characterizing the set of allowed closed curves is an interesting and challenging problem which, very fortunately, has been studied extensively in the mathematical literature \cite{SOCmath}. The allowed curves are called \emph{self-overlapping}. They must satisfy a set of non-trivial, non-local constraints. There are algorithms, based on the idea of Blank cut decomposition, for deciding whether a closed curve is self-overlapping or not. A detailed review of this topic is included in \cite{companion1}.

The most surprising feature revealed by the mathematical analysis is that a given self-overlapping curve may bound several \emph{distinct} distorted disks, the number of which we call the multiplicity of the curve. One can show that all these disks must have the same area, yet the associated metrics are diffeomorphism-inequivalent and thus physically distinct. For instance, the boundary curve depicted in the upper-right corner of Fig.\ \ref{fig1} has multiplicity two; one of the distorted disk it bounds has been explicitly represented, but there is another, see e.g.\ \cite{companion1}. This phenomenon shows that \emph{the degrees of freedom of JT quantum gravity are not in one-to-one correspondence with the shape of the boundary}. Yet, we show in the next section that it is possible to find a boundary description that takes into account the multiplicity.

Putting these ingredients together yields a proposal for a microscopic definition of JT quantum gravity. Let us first discuss the flat case, $\eta=0$. We discretize the closed loops by considering closed polygons on the flat square lattice. Working at fixed boundary length for convenience, the generating function is
\be\label{genW} W_{2n}(t) = \sum_{\alpha\in\text{SOP}_{2n}}\mu_{\alpha}t^{p_{\alpha}} =
\sum_{p\in\mathbb N} W_{2n,p} t^{p},\ee
where $\text{SOP}_{2n}$ is the set of self-overlapping polygons of boundary length $2n$, $p_{\alpha}$ the area (number of lattice squares) of the distorted disk and $\mu_{\alpha}$ its multiplicity. One can show \cite{companion1} that, when $t<1$, the model is in a deflated, branched-polymer-like phase, with $\ln W_{2n}(t)$ proportional to $n$, and that when $t>1$, we have an inflated phase with $\ln W_{2n}(t) \propto n^{2}\ln t$. This is reminiscent of other interesting random path models, as the self-avoiding polygon model, see e.g.\ \cite{SAPtextbook}. We conjecture that the critical point $t=1$ can be used to define a continuum limit. Introducing a cut-off $\ell_{0}$, with lattice squares area $\ell_{0}^{2}$, the continuum limit is $n\rightarrow\infty\, ,\ \ell_{0}\rightarrow 0\, ,\ 2n\ell_{0}^{1/\nu}=\beta\ \text{fixed}$, for a critical exponent $\nu$ that must be such that $\frac{1}{2}\leq\nu\leq 1$. The Hausdorff fractal dimension of the boundary curves is $1/\nu$. The numbers $w_{2n;r}=\sum_{p}p^{r}W_{2n,p}$ are conjectured to scale as $\smash{g_{*}^{-2n}(2n)^{2\nu(r-\vartheta)-1}}$, for a non-universal ``connective constant'' $1/g_{*}\geq 1$ and another universal exponent $\vartheta$. We predict in the next section that $\nu=1/2$ and $\vartheta=2$, to be compared to the random walk values $\nu=1/2$, $\vartheta =1$ or to the self-avoiding polygon values $\nu=3/4$, $\vartheta = 1$. Moreover, in the limit $t\rightarrow 1^{-}$, $\ell_{0}\rightarrow 0$, $(1-t)/\ell_{0}^{2} = \La/(16\pi)$ fixed, we conjecture that the generating function scales as 
\be\label{scalingform} W_{2n}(t)\sim e^{-\beta\la}\beta^{-1-2\nu\vartheta}f(\La\beta^{2\nu}) = \mathscr Z\ee
and matches with the JT partition function \eqref{ZJTgen}. The analysis of the next section actually suggests that the scaling laws may get logarithmic corrections, in a way similar to the $c=1$ matrix quantum mechanics \cite{MQMlog}. The construction can be generalized to the cases of negative or positive curvatures \cite{companion1}, in the same way as one can define random walks in hyperbolic space or on the sphere, see e.g.\ \cite{hyperbrw}. The fact that the boundary curves are fractal, of dimension $1/\nu>1$, is a short-distance effect and will thus be equally valid for zero, negative or positive curvature. This implies that the naive boundary length $\ell$ is replaced in all the models by a renormalized parameter $\beta$, with $\beta^{\nu}$ having the dimension of length. Note that approaches based on the reparameterization ansatz deal with smooth boundary curves and are thus at best effective long-distance descriptions.
 
A subtle question concerns the existence of the models for negative cosmological constant. A necessary condition is the existence of all the moments $\langle A^{k}\rangle$ of the area $A=p\ell_{0}^{2}$ at $\La=0$. We conjecture that the negative and zero curvature models exist for negative cosmological constants (up to some strictly negative critical value for the flat model) and that the scaling function $f$ appearing in Eq.\ \eqref{scalingform} can be analytically continued to negative values of its arguments in these cases. The situation for the positive curvature model seems to be much less favorable \cite{companion1}.

\section{Continuum approach and $\text{CFT}_{1}$}

\subsection{General discussion}

The continuum approach is based on the following observations. Acting with diffeomorphisms, we can put any metric on $\disk$ in the conformal gauge $g = e^{2\Sigma}\delta_{0}$, where $\delta_{0} = |\d z|^{2}=\d\rho^{2}+\rho^{2}\d\theta^{2}$ is the round flat metric of area $\pi$. We call $\Sigma$ the Liouville field. The constraint of constant curvature $R=2\eta$ is equivalent to $\Delta_{0}\Sigma = \eta e^{2\Sigma}$, where $\Delta_{0}$ is the positive Laplacian for the metric $\delta_{0}$. We then rely on the following theorem (and its generalization to the case of distribution-valued boundary fields, see below). \emph{Let $\sigma:\Sone\rightarrow\mathbb R$ be a continuous function defined on the boundary of the disk. Then, when $\eta=0$ or $\eta=-1$, there exists a unique solution $\Sigma_{\sigma}\in C^{\infty}(\disk)$ of $\Delta_{0}\Sigma = \eta e^{2\Sigma}$ such that $\Sigma_{\sigma|\partial\disk}=\sigma$.} When $\eta=0$, $\Sigma_{\sigma}$ is a harmonic function expressed in terms of $\sigma$ by using the Poisson kernel. If we use the Fourier decomposition,
\be\label{SigsigFourier} \Sigma_{\sigma} = \sum_{n\in\mathbb Z}\sigma_{n}\rho^{|n|}e^{in\theta}\quad \text{if}\quad \sigma = \sum_{n\in\mathbb Z}\sigma_{n}e^{in\theta}\, .\ee
When $\eta=- 1$, the theorem is still valid \cite{Lth} even though an explicit formula for $\Sigma_{\sigma}$ no longer exists. \emph{In these cases, the degrees of freedom of JT gravity are thus encoded in the boundary Liouville field $\sigma$, modulo the action of the group $\PslR$ of disk automorphisms.} When $\eta=+1$, the situation is more subtle and will be discussed elsewhere \cite{companion2}. Let us simply note here that these subtleties do not affect the UV structure.

One can reconstruct the distorted disk from $\sigma$. In the flat case, the immersion $F_{\sigma}$ associated with $\sigma$ is such that $F_{\sigma}'=e^{H_{\sigma}}$, where $H_{\sigma}$ is a holomorphic function whose real part is $\Sigma_{\sigma}$. The associated self-overlapping boundary is $\alpha_{\sigma}:\theta\mapsto F_{\sigma}(e^{i\theta})$. This is how the plots in the lower insets of Fig.\ \ref{fig1} are obtained, starting with typical realizations of $\sigma$ according to the probability law constructed from the action \eqref{HilbLiouaction} below. The description in terms of $\sigma$ automatically takes into account the multiplicity. If a given self-overlapping boundary has multiplicity $p$, then there will be $p$ $\PslR$-inequivalent boundary fields $\sigma_{k}$ that all yield the same immersed boundary, $\im\alpha_{\sigma_{1}}= \cdots = \im\alpha_{\sigma_{p}}$, but have different interiors.

The construction of quantum JT gravity thus boils down to the construction of $\PslR$-invariant probability measures on the space of boundary Liouville fields. Note that neither this space nor the action of $\PslR$ depend on $\eta$ and thus, abstractly, the construction of the models does not depend on $\eta$! Of course, the formulas expressing the interesting observables, like the area, will depend on $\eta$. A nice illustration of this principle is as follows. Assume that we are looking for a probability measure such that $\sigma$ is continuous almost surely. In this case, the boundary is smooth and its length $\ell = \int_{0}^{2\pi}\!e^{\sigma}\d\theta$ is well-defined. Expressing $e^{\sigma} = \frac{\ell}{2\pi}(f^{-1})'$, where $f\in\diffSp$ is a diffeomorphism of the circle, one can check that $\PslR$ simply acts by left-multiplication. Any $\PslR$ left-invariant measure on $\diffSp$ thus defines a model. One may choose to use the Malliavin-Shavgulidze measure \cite{MSmeasure}, which is nothing but the Schwarzian model used in the physics literature. The surprise is that this works just as well in negative, zero or positive curvature! Of course, the diffeomorphism $f$, which parameterizes the most general distorted disks, does not match with the diffeomorphism used in the reparameterization ansatz, which allows one to describe only a very small subset of these disks; but one can show that they do match at leading order in the near AdS limit \cite{companion2}.

%Our claim the the correct degree of freedom in JT gravity is the boundary Liouville field is absolutely central to our work but contradicts claims made in the literature, so let us briefly clarify some important and potentially confusing points. It is very important to understand that, for any two-dimensional quantum gravity theory on the disk, \emph{the Liouville field $\Sigma$ has free boundary conditions}. Imposing a priori a contraint on the boundary for $\Sigma$ would artificually, and incorrectly, restrict the space of metrics. In particular, imposing a Dirichlet b.c.\ would reduce the space of constant curvature metrics to a point, as is clear from our discussion above. On the other hand, the normal derivative of $\Sigma$ is related to the extrinsic curvature $k$ of the boundary according to 

\subsection{Boundary $\text{CFT}_{1}$}

$\PslR$-invariant measures on the space of boundary Liouville fields are far from being unique. The physically relevant measure is the continuum limit of the SOP model introduced in Section \ref{SOPSec}, or the measure relevant for a quantum description (we conjecture that these two criteria are actually equivalent). Our goal is now to construct this measure, starting from the usual path integral with the formal ultralocal measure, adapting to the case of JT the arguments used sucessfully in the past for Liouville. We actually consider a generalization of pure JT, coupling an arbitrary bulk matter CFT of central charge $c$ to gravity. We are going to argue that the correct measures, which depend on $c$, are such that $\sigma$ is almost surely infinite at any point and must thus be treated as a random distribution. As a consequence, the boundary self-overlapping curves are fractals and the length $\ell$ is infinite.

An essential point to understand for correct path integral treatment is that the Liouville field $\Sigma$ has free boundary conditions. Imposing any kind of boundary constraint on $\Sigma$ would incorrectly restrict the space of metrics over which we want to integrate. We separate the bulk and the boundary degrees of freedom by writing $\Sigma = \Sigma_{\text B} + \Sigma_{\sigma}$ where, as explained above, $\Sigma_{\sigma}$ is determined by the boundary field $\sigma$ and $\Sigma_{\text B}$ satisfies Dirichlet boundary conditions, $\Sigma_{\text B|\partial\disk} = 0$. In JT gravity, there is a $\delta$-function in the path integral, coming from integrating out the dilaton field $\Phi$, enforcing the constraint of constant bulk curvature $R=2\eta$. It is straightforward to check that this allows to perform explicitly the integral over $\Sigma_{\text B}$, yielding a partition function of the form
\begin{multline}\label{Zformal1} \mathscr Z = \int D\sigma \prod_{k=1}^{3}\delta\Bigl(\int_{\frac{2\pi}{3}(k-1)}^{\frac{2\pi}{3}k}e^{\sigma}\d\theta - \frac{\ell}{3}\Bigr) e^{\sigma(\frac{2\pi}{3}) + \sigma(\frac{4\pi}{3})}\\|\det (\Delta_{\sigma} - 2\eta)|^{-1} e^{-\frac{\La}{16\pi}A - \frac{26-c}{24\pi}S_{\text L}[\Sigma_{\sigma}]}\, .
\end{multline}
The $\delta$-functions in the first line implement the boundary length constraint and convenient gauge fixing conditions for the $\PslR$ disk automorphisms, with associated Fadeev-Popov determinant given by \smash{$e^{\sigma(\frac{2\pi}{3}) + \sigma(\frac{4\pi}{3})}$}. The determinant in the second line must be computed with Dirichlet boundary conditions and comes from treating the $R=2\eta$ constraint. The action $S_{\text L}$ is the standard Liouville action, with the familiar $\frac{26-c}{24\pi}$ factor produced by the ghosts plus matter CFTs.

For conciseness, let us now restrict ourselves to the flat $\eta=0$ case. The determinant in \eqref{Zformal1} is equivalent to the contribution of a $c=2$ CFT, which may be understood as coming from the original dilaton field plus the familiar conformal anomaly associated with the bulk Liouville field itself. Overall, the determinant is thus absorbed by replacing $26-c$ by $24-c$ in front of the Liouville action. Rescaling $\Sigma_{\sigma} = \phi/Q$ with $Q = \sqrt{(24-c)/6}$ and writing $\phi_{|\partial\disk}=\varphi$, one finds that the action reduces to a nice non-local $\PslR$-invariant ``Hilbert-Liouville'' action for $\varphi$,
\be\label{HilbLiouaction} S = \frac{1}{4\pi}\int_{0}^{2\pi}\bigl(\varphi\Hilb[\varphi'] + 2Q\varphi\bigr)\d\theta = \sum_{n\geq 1}n|\varphi_{n}|^{2} + Q\varphi_{0}\, ,
\ee
where $\Hilb$ denotes the Hilbert transform and the $\varphi_{n}$ are the Fourier coefficients of $\varphi$. This action implies that a typical boundary field has $\varphi_{n} = Q\sigma_{n}$ of order $1/\sqrt{n}$, and thus diverges pointwise everywhere almost surely \cite{RandomFourier}. The field $\varphi$ is a random distribution, but a fundamental point is that the field $\phi$ in the interior is still uniquely determined by $\varphi$, according to \eqref{SigsigFourier}, and is smooth. Writing $\varphi = \varphi_{0} + \check\varphi$, the action \eqref{HilbLiouaction} implies that $\check\varphi$ is a one-dimension log-correlated Gaussian free field, matching with the boundary field in ordinary Liouville theory. However, to the contrary of ordinary Liouville, its extension in the interior is smooth, with a Green function $G_{\check\phi\check\phi}(z_{1},z_{2}) = -2\ln|1-\bar z_{1}z_{2}|$ that is regular at coinciding bulk points. 

The above analysis implies that tools from Liouville theory can be straightforwardly adapted to JT gravity. Exactly as in Liouville, the correct boundary length operator is renormalized and is given by $\beta =\int_{0}^{2\pi}\!\!:\!e^{\frac{\gamma}{2}\varphi}\!:\d\theta$, with $Q = \frac{2}{\gamma} + \frac{\gamma}{2}$ and $0\leq\gamma\leq 2$. However, in sharp contrast with Liouville, the bulk area operator is not renormalized and is given by $A=\int_{\disk}e^{2\phi/Q}\d^{2}z$. This is of course the consequence of the fact that the bulk geometry does not fluctuate in JT gravity. These remarks allow one to write the correct, rigorously defined partition function, by replacing the boundary conformal factor $e^{\sigma}$  by its renormalized version $:\! e^{\frac{\gamma}{2}\varphi}\!:$ in Eq.\ \eqref{Zformal1}. A direct argument \`a la DDK \cite{Liouvilleref1} then yields a partition function of the form of Eq.\ \eqref{scalingform}, with
\be\label{critexponents} \nu = \frac{1}{2}\Biggl[1+\sqrt{\frac{c}{c-24}}\Biggr]\, ,\quad \vartheta = 2 - \frac{c}{12}\, \cdotp\ee
The pure gravity $c=0$ case corresponds to $\nu=1/2$ and $\vartheta =2$, and we have a barrier at $c=0$, alike the $c=1$ barrier in ordinary Liouville. The limit $c\rightarrow -\infty$ is semiclassical, with $\nu\rightarrow 1^{-}$, as expected \cite{more1}. At $c>0$, we conjecture that the geometry is made of multiple ``baby universes'' connected by narrow wormholes, as suggested by the numerics, see Fig.\ \ref{fig1}.

\section{Outlook}

We have explained that the correct fluctuating boundary picture for finite cut-off JT gravity is a new model of random self-overlapping curves that must be counted with a non-trivial multiplicity. This implies that many ideas about JT gravity must be seriously amended. The degrees of freedom are not in bijection with the shape of the boundary. The gauge-theoretic BF approach \cite{BFpapers} will not work, because in this approach the metric is not necessarily positive-definite, although this constraint is essential to restrict the configuration space from  arbitrary random loops to self-overlapping loops. The trumpet decomposition, which has been instrumental in the study of higher topologies \cite{SSS}, is not valid, because, due to wildly fluctuating boundaries, typical geometries do not have geodesics along which the gluing of the trumpet can be made \cite{companion1}.

Fortunately, powerful tools developed for Liouville can be imported to JT gravity. The microscopic structure is described in terms of the log-correlated boundary Liouville field. This makes it plausible that an exact solution can be found, at least in the flat case, along the lines of DOZZ and FZZT \cite{Liouvilleref2}. A direct combinatorial approach for the self-overlapping polygon model is also worth pursuing, both for its own intrinsic interest (it carries new very interesting questions, e.g.\ one would like to assess the role of the curves with non-trivial multiplicity), and for comparison with the continuum approach. We note that a matrix model formulation can be built \cite{companion1}, using the idea of dually weighted graphs \cite{dualgraph}.

Our work has focused on the Euclidean theories, but the ideas can be extended to real-time physics, yielding explicit models of Lorentzian random geometries. The degrees of freedom are encoded in a distribution-valued Liouville field defined on a timelike (for negative or zero curvature) or spacelike (for positive curvature or closed universes) boundary, from which smooth notions of time and space emerge. The relation to canonical quantization and the consequences for the operator algebra of observables and holography, including in a cosmological context, will be explored in the future. One can also speculate on the new critical string theory at $c=24$ and the supersymmetric generalizations.

\section*{Acknowledgements}

This work has a very long and tortuous history, tightly intertwined with the sad period of the Covid pandemic. Some results, including the role played by self-overlapping curves in JT gravity, were sketched as early as 2021 during a seminar at the Institute for Advanced Study in Dublin, Ireland, and at the workshop on Quantum Geometry, Field Theory and Gravity in Corfu, Greece. I am particularly grateful to the organisers of the Quantum Gravity, Random Geometry and Holography programme at the Institut Henri Poincar\'e in Paris, France (J.~Barrett, D.~Benedetti, J.~Ben Geloun and R.~Loll, January and February 2023), of the workshop on Random Geometry in Math and Physics in Nijmegen, The Netherlands (T.~Budd, March 2023) and to the APCTP in Pohang, South Korea (J.~Yoon, April and May 2023), for providing me with an exceptional scientific and working environment, which has enabled the work presented here to mature.

This work is supported in part by the International Solvay Institutes.

\end{document}